# Energy Absorption and Low Velocity Impact Response of Open-Cell Polyurea Foams

Brian J. Ramirez [1], and Vijay Gupta[1, *]

[1]Department of Mechanical and Aerospace Engineering, University of California, Los Angeles, California 90095, USA; bjramirez@ucla.edu

*Correspondence: vgupta@ucla.edu; Tel.: +1-310-825-0223

## Abstract

The energy absorption and impact attenuation of low density polyurea (PU) foams (140 - 220 kg/m$^3$) is presented. The stress-strain behavior, energy absorption, cushioning efficiency, and energy return (resilience) are measured at a quasi-static strain rate using an Instron load frame. In addition, the low velocity impact attenuation of the polyurea foams at 5 J and 7 J impact energies were investigated according to the American Society for Testing and Materials (ASTM) test methods F1976. The polyurea foams were then compared with widely used ethylene vinyl acetate (EVA) and thermoplastic polyurethane (TPU) foam technologies at similar densities. The PU foams displayed a 5 % increase in cushioning efficiency and 30 % greater energy absorption over TPU and EVA foams under quasi-static compression. Under impact testing, PU foams resulted in a reduction of up to 28.6 % and 36.9 % in peak transmitted impact forces when compared to EVA and TPU foams of the same thickness (20 mm) at 5 J and 7 J, respectively. These properties should allow these materials to have a wide range of cushioning/impact applications, especially in protective body and headgear systems.

## 1. Introduction

Elastomeric foams are widely used for impact management in several mechanical, civil, and aerospace structures as well as in personal protective gears [1,2]. Foams dissipate impact energy through local viscoelastic processes through bending, twisting, and buckling of the struts that form the skeleton of its structure [1-3]. The compressive stress-strain behavior of a typical viscoelastic foam exhibits three distinct regions: (i) a linear elastic region associated with cell wall bending, (ii) a stress plateau region that results from continuous collapse of the foam's cell structure on itself through non-linear fully recoverable (viscoelastic foams) buckling of its struts, and (iii) a densification region where the already-collapsed cells begin to compress against each other to near full density, resulting in a steeply rising stress. The amount of energy absorbed or dissipated by the foam is roughly equal to the area under its stress-strain curve [4-6].

Cushioning and impact attenuation in particular, are matters of energy management. The goal is to accommodate the impact energy while maintaining the load at or below a (acceptably low) threshold level. A superior cushioning material or system is one that that absorbs more energy at the same load, or absorbs the same energy at a lower load. Today, foams are used in practically every industry. Despite their ubiquity and maturity, there are still areas where improvements in the foam technology can be made, especially those that pertain to impact applications [2,7].

Motivated by these goals, in previous communications, we presented a polyurea (PU) based viscoelastic foam that became increasingly more efficient at absorbing energy than plastic type foams under high strain rates, similar to those found in ballistic applications (> 2000 $s^{-1}$) [8]. The combined benefits of temperature stability and rate sensitivity of low density ($\leq$ 300 kg/m$^3$) viscoelastic PU foams were demonstrated by integrating them as liners into football helmets, and evaluating their performance using the NOCSAE (National Operating Committee for Standards in Athletic Equipment) football helmet standard [9]. With a glass transition temperature ($T_g$) of -50 °C, they were also able to maintain their ambient impact properties at low (-15 °C) and high temperatures (50 °C). Lastly, a composite foam concept that involved infiltrating a polyurea-based foam through an open 3D elastic lattice structure that allows several energy absorbing mechanisms to operate synergistically and sequentially at varying length scales, was introduced [10].

The purpose of this paper is to investigate the cushioning and impact attenuation properties of low density open-cell polyurea foams, according to the American Society for Testing and Materials (ASTM) test methods F1976 [11]. The polyurea foams were then compared with commercially available ethylene-vinyl acetate (EVA) and thermoplastic polyurethane (TPU) foams, which are widely used as liners in protective body and headgear systems [1,2,7].

## 2. Materials and Methods

### 2.1. Samples and Materials

Open-cell polyurea foams were fabricated by mixing modified Methylene Diphenyl Diisocyanate compound (Isonate 143L MDI; Dow Chemical) with oligomeric diamine (Versalink P1000; Air Products) with a 1:4 ratio. Polyureas of this stoichiometry are known to have behaviors within the viscoelastic regime [12-14]. The mixture was stirred using a high rpm electric stirrer to introduce small air bubbles to create froth. Next, the PU mix was poured into a 90 mm x 90 mm x 90 mm mold and placed inside a vacuum oven that was preheated to 80 °C. The oven was slowly evacuated until a desired foam height was reached. The vacuum was then maintained at that pressure for 10 minutes so that it could support the cellular structure of the foam during curing. The density of the foam was controlled by adjusting the amount of air and frothing.

Foam core samples were cut from the center of the manufactured blocks, free of voids, defects, and top/bottom skins. The dimensions and weight of the samples were measured within 1 %, as indicated in ASTM F1976. Sample densities of $140 \pm 1.4$, $170 \pm 1.6$, $200 \pm 1.1$, and $220 \pm 1.2$ kg/m$^3$ were prepared using the same polyurea mix with a density of 1070 kg/m$^3$. Hereinafter these samples are referred to as PU140, PU170, PU200, and PU220, respectively. These densities were chosen because they fall within the range used in athletic footwear and body protection equipment [1,2,7]. For comparing the performance of the PU foams, ethylene-vinyl acetate (EVA) and thermoplastic polyurethane (TPU) foams were chosen as reference foams which are widely used in athletic footwear as well as in body/head protective equipment. EVA samples with densities of $170 \pm 0.5$ kg/m$^3$ (EVA170) and $220 \pm 0.5$ kg/m$^3$ (EVA220) along with TPU samples with densities of $170 \pm 0.5$ kg/m$^3$ (TPU170) and $220 \pm 0.5$ kg/m$^3$ (TPU220) were obtained commercially.

Figure 1 shows the micrographs of these foam samples examined under a scanning electron microscope (SEM) (Nova Nano 230) by using its low vacuum secondary detector (LVD) at a voltage of 5 kV. This minimized the charging in our soft polymeric samples and allowed imaging without the need of a conductive coating on its surface. The PU foams consist of open spheroid cells with wall apertures (perforations) as shown in Figure 1a. The average cell sizes are $1000 \pm 258$ μm, $600 \pm 153$ μm, $400 \pm 72$ μm, and $300 \pm 60$ μm for PU140, PU170, PU200, and PU220, respectively. Figures 1b and 1c show the SEM micrographs of TPU220 and EVA220 which are comprised of irregular polyhedral closed cells with an average cell size of $130 \pm 72$ μm and $160 \pm 66$ μm, respectively.

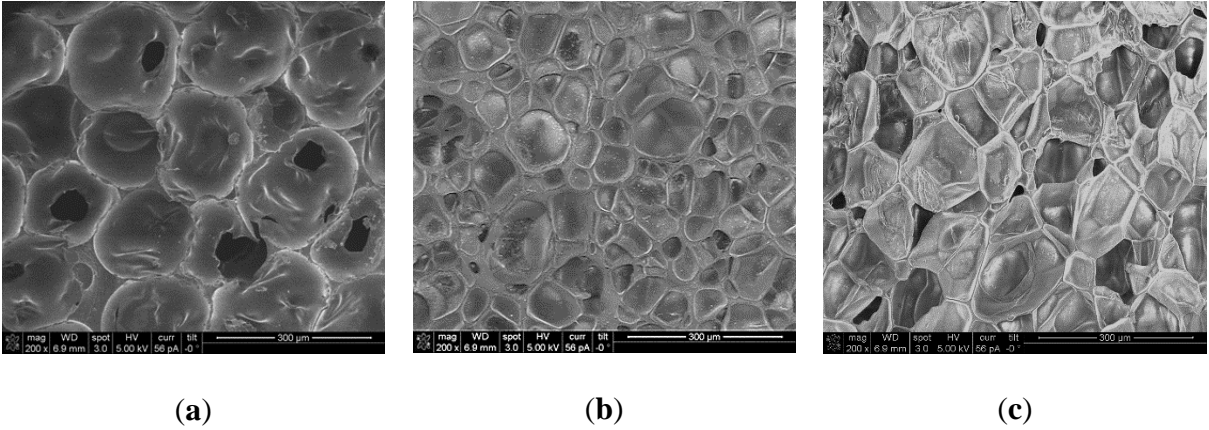

(**a**)                   (**b**)                  (**c**)

**Figure 1.** SEM images of: (**a**) PU220; (**b**) TPU220; (**c**) EVA220. The black spots in (**a**) are apertures (or perforations) in the cell walls, while in (**c**) they are voids within the closed cellular structure.

## 2.2. Quasi-static Compression Tests

Low strain rate ($\leq 10^{-1}$ s$^{-1}$) compression tests were conducted using an Instron Micro-Tester (Model 5942) equipped with a 2 kN load frame. For these tests, disc samples of 28 mm nominal diameter and 10 mm thickness were cut from a large molded block. Samples were then compressed under stroke-control with a constant cross-head velocity of 24 mm/min. This yielded a constant strain rate of 0.04 s$^{-1}$. Force-displacement data obtained from the machine was converted into the stress-strain data using the sample dimensions. For each density, three specimens were tested. The plots presented here represent the average of data obtained from these experiments.

## 2.3. Impact Tests

Impact tests were carried out using an Instron DynaTup (Model 8250) drop-weight impact tester. Specimens of 75 x75 x 20 mm sections were placed on a 45 mm-diameter flat force plate and impacted using an 8.5 kg balanced mass attached to a 75 mm-diameter flat stainless steel indenter head, as per ASTM F1976 standards. Impact tests were carried out at 5 J and 7 J energies under ambient (23 °C) conditions. The impact energy was varied by changing the drop height of the indenter head. The transmitted force was measured using a force transducer (Kistler Instruments, Model 9041A) with a full scale output of 90 kN, sensitivity of 4.3 pC/N, and a threshold < 0.01 N. The displacement of the sample was measured by tracking the impactor as it contacts the sample during impact using high speed photography (Olympus i-Speed 3 camera with capabilities of 150,000 fps and a recording time of 5 seconds). For each impact test, three specimens were tested. The plots presented here represent the average of data obtained from these experiments. A schematic of the experimental set-up is shown in Figure 2.

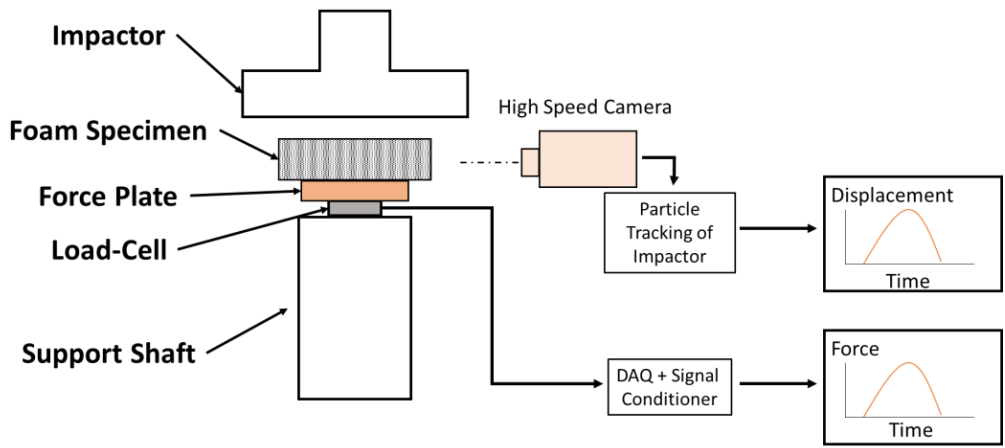

**Figure 2**. Experimental set-up of the low velocity impact tests.

*2.4. Reduction of Data: Cushioning Efficiency and Impact Attenuation*

Cushioning and impact attenuation is an energy management problem. The goal is to accommodate an impact, quantifiable as an energy input, in a manner that minimize the consequent loads and maintains them below a "safe" threshold level. Thus, the properties of cushioning materials can be described in terms of: (i) Cushioning capacity: an index of the amount of energy that a cushioning material or system can usefully accommodate and is often expressed as the energy capacity per unit volume (energy absorbtion-*W*) of a material and (ii) Cushioning performance: a measure of the load outcomes produced by an impact, expressed in terms of absolute force, pressure or acceleration. The cushioning capacity and performance of *real* cushioning systems are finite and constrained. All real materials have a are finite in thickness and have a useful compression limit that is somewhat less than that thickness. At compression levels beyond the useful limit, the material "bottoms out" and the stiffness of the system approaches infinity.

To assist in this selection process, the energy absorption (*W)* for various foams is plotted as a function of the stress so that the user can select foam with the maximum energy absorption capabilities for the selected threshold stress. The energy absorption in this paper was determined by calculating the area between the loading and unloading stress-strain curves. Another way to assist in this selection process is to plot the cushioning efficiency (*CE*) of the foam, defined as the energy-load ratio, $W/\sigma$. For a cushioning material, when expressed in geometry independent terms (strain energy density per unit stress) the ratio varies between zero and one. Efficiency generally increases up to the onset of densification, then declines at higher strains. Therefore, it can approximate the densification strain, the limit of energy absorbtion (*W)*, and the limit of the threshold stress. The optimal material will generally be one that has the highest cushioning efficiency in the range of loads and impact energies it will experience in the application. It is important to measure *W* and *CE* under the prevailing temperature and strain rate conditions in service as both of these parameters greatly influence the underlying stress-strain characteristic of the foam material.

Furthermore, the outcome of an impact test is the peak impact shock (force) or g-max score (acceleration, g). Lower peak impact values indicate less impact shock and hence *more* impact attenuation. In addition, the impact cushioning effiency can also be calculated by taking the ratio of the theoretically minimum g-max to the g-max score measured under identical impact conditions. The theoretically minimum possible g-max score for an ideal cushioning system is the impact energy divided by the thickness of the material and the mass of the impactor. Another impact parameter of interest is the energy return of the foam during impact, the percentage of the input energy that is returned and not dissipated. This is calcuated by taking the ratio of the area under the unloading part of the force-displacement curve obtained during the impact and divide it by the area under the loading curve. Higher values indicate greater "rebound" under the specific impact conditions. This parameter is a good indication of how resilient (or responsive) the foam system is. Lastly, the peak deflection of the foam during impact is also recorded as it is an indication of how compliant the foam is and how much design space is required when using this foam. The optimal foam under identical impact condictions is therefore one which minimizes the peak shock loading (higest impact efficiency), while exhibiting the highest energy return and the least compression.

In this paper we present the stress-strain characteristics, cushioning parameters (cushioning efficiency as a function of stress, strain, and energy absorption), and impact properties (peak transmitted force and impact cushioning efficiency) for the PU foam material and compare their performance with widely used EVA and TPU foams.

## 3. Results and Discussion

The variation in the experimental data shown in Figures 3 - 7 range from 0.3 - 11 %. The one standard deviation for the properties of interest are shown in Tables 1 - 4.

*3.1. Quasi-static Compression Test Results*

Figure 3 shows the compressive stress-strain behavior of the PU, TPU, and EVA foam materials at a quasi-static (0.04 s$^{-1}$) strain rate. The compressive stress-strain behavior of all foam materials exhibit the classical linear, plateau, and densification regimes. In addition, the foams display an increase in the Young's modulus, increase in plateau stress, and decrease in the densification strain with increasing density. Table 1 summarizes the quasi-static (0.04 s$^{-1}$) compressive stress-strain characteristic of all foams.

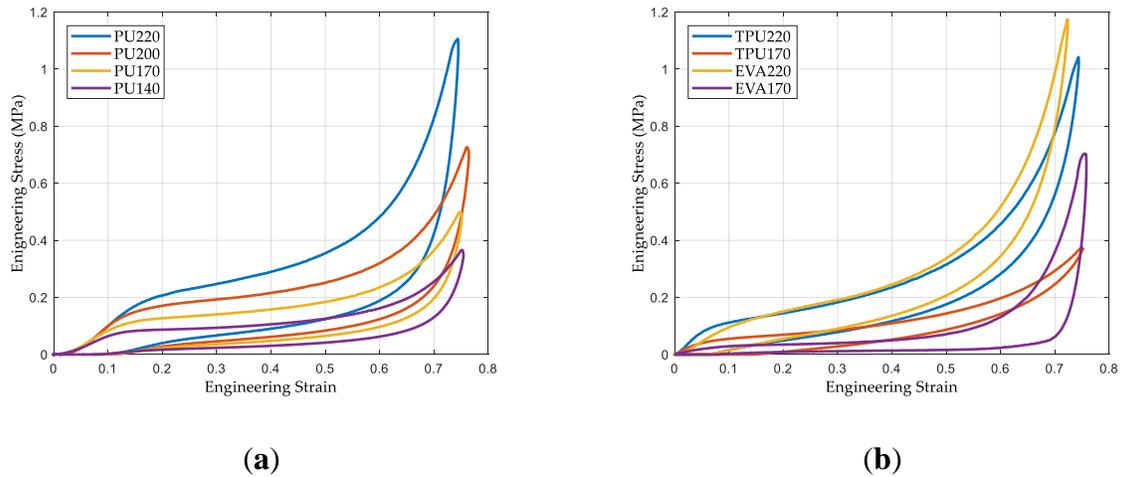

(**a**) (**b**)

**Figure 3.** Quasi-static (0.04 s$^{-1}$) compressive stress-strain behavior of: (**a**) PU foams; (**b**) TPU and EVA foam systems.

**Table 1.** Quasi-static (0.04 s$^{-1}$) Compressive Material Characterization Test Results

| Material | Density (kg/m³) | Elastic Moduli | | Plateau Onset | |
|---|---|---|---|---|---|
| | | Initial (MPa) | Plateau (MPa) | Strain | Stress (MPa) |
| PU140 | 140 | 0.88 ± 0.007 | 0.12 ± 0.005 | 0.12 ± 0.006 | 0.08 ± 0.003 |
| PU170 | 170 | 1.17 ± 0.009 | 0.17 ± 0.007 | 0.14 ± 0.003 | 0.11 ± 0.001 |
| PU200 | 200 | 1.52 ± 0.011 | 0.22 ± 0.002 | 0.17 ± 0.004 | 0.16 ± 0.005 |
| PU220 | 220 | 1.66 ± 0.012 | 0.42 ± 0.002 | 0.19 ± 0.008 | 0.20 ± 0.007 |
| TPU170 | 170 | 0.99 ± 0.003 | 0.18 ± 0.002 | 0.09 ± 0.007 | 0.05 ± 0.001 |
| TPU220 | 220 | 1.94 ± 0.012 | 0.38 ± 0.004 | 0.10 ± 0.004 | 0.1 ± 0.002 |
| EVA170 | 170 | 0.32 ± 0.002 | 0.07 ± 0.001 | 0.09 ± 0.003 | 0.03 ± 0.001 |
| EVA220 | 220 | 1.20 ± 0.016 | 0.41 ± 0.002 | 0.14 ± 0.007 | 0.12 ± 0.006 |

Figure 4 shows the quasi-static (0.04 s$^{-1}$) energy absorption and cushioning efficiency properties for the PU foam samples, while Figure 5 shows the equivalent properties for TPU and EVA foam systems. The cushioning efficiency is 34 % for PU and 29 % for EVA and TPU foams at similar values of stress and strain. It can also be noted that the stress for maximum efficiency increases non-linearly with foam density for all materials. Although, PU, TPU, and EVA foam samples display similar densification strain values of 0.60, PU foams have up to 30 % greater energy absorption capabilities. This is attributed to the higher elastic moduli and plateau stress of the PU foams. Lastly, the energy return of the PU foams is about 38 %, thus, they can be categorized as moderately resilient foams, meaning they are more "bouncy" or responsive to indentation loads than "memory" type foams. Table 2 summarizes the quasi-static compressive cushioning properties for each foam material.

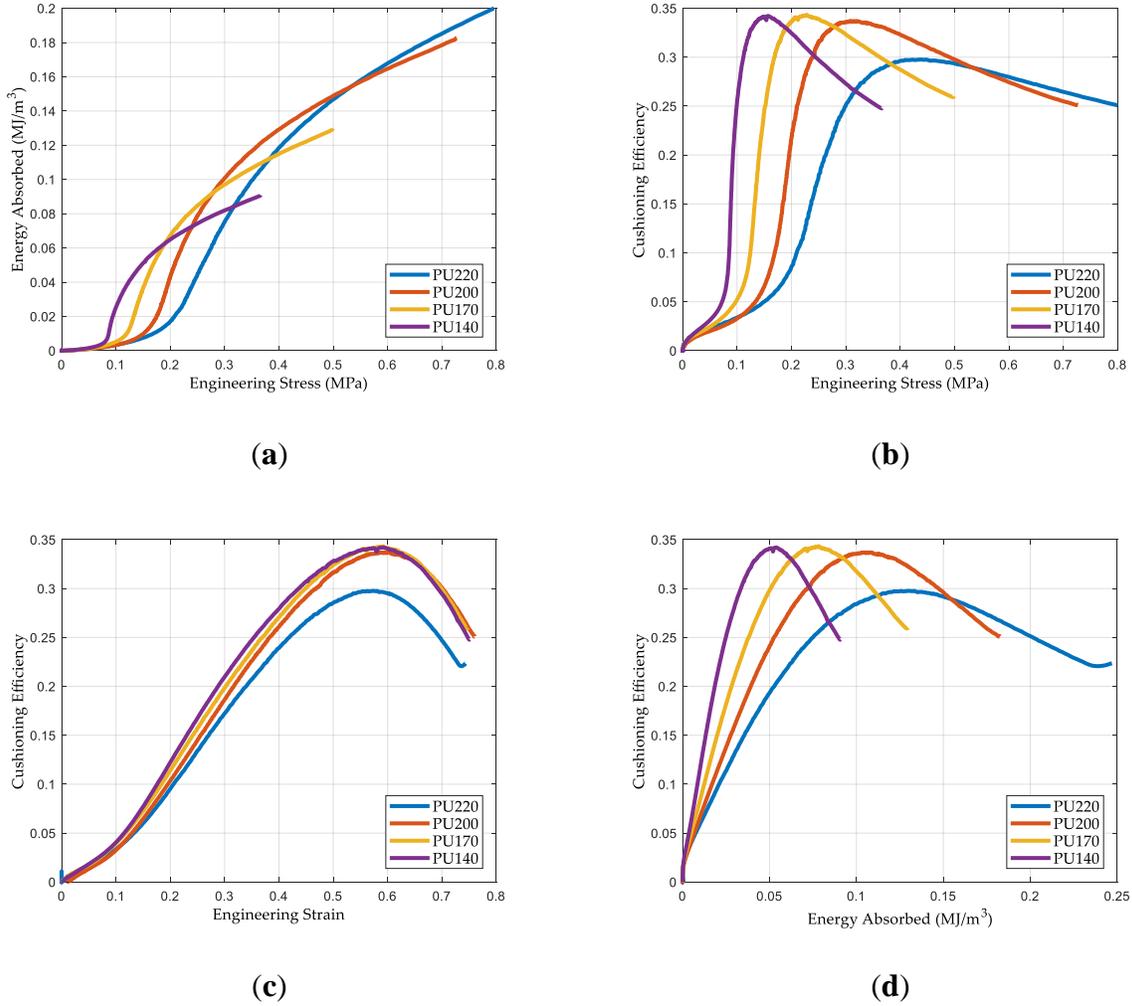

**Figure 4.** Quasi-static (0.04 s$^{-1}$) cushioning properties of PU foams: (**a**) Energy absorption as a function of stress; (**b**) Cushioning efficiency as a function of stress; (**c**) Cushioning efficiency as a function of strain; (**d**) Cushioning efficiency as a function of energy absorbed.

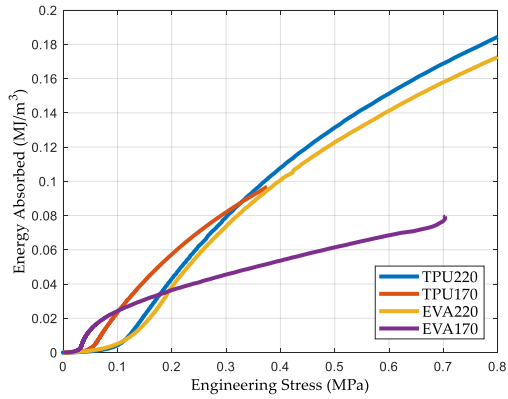
(**a**)

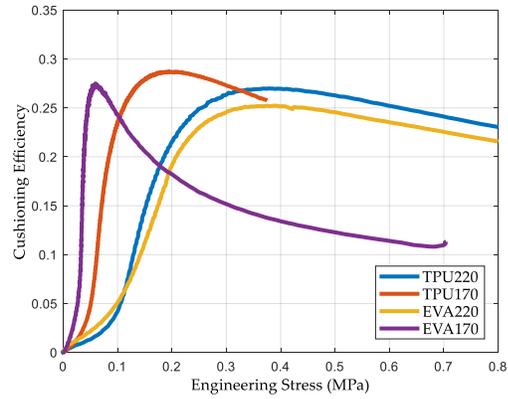
(**b**)

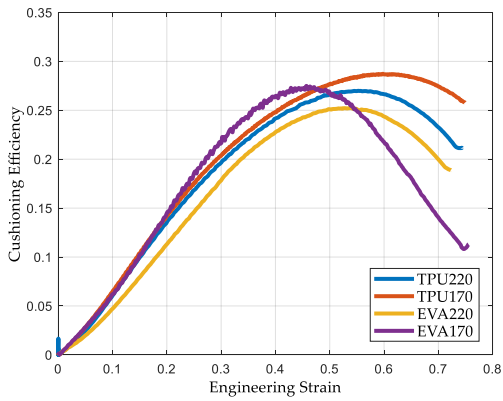
(**c**)

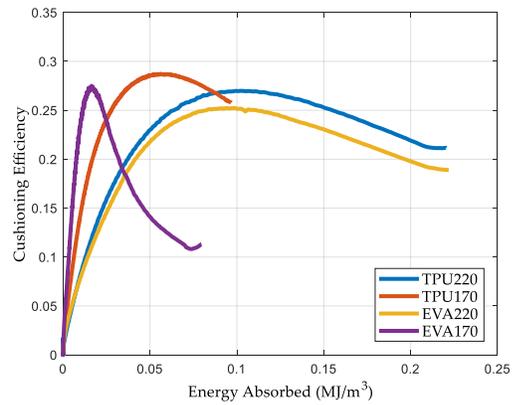
(**d**)

**Figure 5.** Quasi-static (0.04 s$^{-1}$) cushioning properties of TPU and EVA foams: (**a**) Energy absorption as a function of stress; (**b**) Cushioning efficiency as a function of stress; (**c**) Cushioning efficiency as a function of strain; (**d**) Cushioning efficiency as a function of energy absorbed.

**Table 2.** Quasi-static (0.04 s$^{-1}$) Cushioning Efficiency and Energy Return Results

| Material | Density (kg/m$^3$) | Peak Cushioning Efficiency | | | | Energy Return (%) |
|---|---|---|---|---|---|---|
| | | Value | Strain | Stress (MPa) | Energy Absorbed (MJ/m$^3$) | |
| PU140 | 140 | 0.34 ± 0.002 | 0.60 ± 0.005 | 0.16 ± 0.004 | 0.05 ± .003 | 36.61 ± 1.52 |
| PU170 | 170 | 0.34 ± 0.001 | 0.59 ± 0.003 | 0.22 ± 0.003 | 0.08 ± .002 | 36.62 ± 2.36 |
| PU200 | 200 | 0.34 ± 0.003 | 0.59 ± 0.007 | 0.32 ± 0.002 | 0.12 ± .005 | 37.57 ± 1.71 |
| PU220 | 220 | 0.30 ± 0.005 | 0.57 ± 0.005 | 0.44 ± 0.002 | 0.13 ± .004 | 38.81 ± 1.42 |
| TPU170 | 170 | 0.29 ± 0.002 | 0.59 ± 0.003 | 0.20 ± 0.007 | 0.06 ± 0.003 | 57.79 ± 2.26 |
| TPU220 | 220 | 0.27 ± 0.003 | 0.55 ± 0.006 | 0.38 ± 0.004 | 0.10 ± 0.005 | 59.23 ± 1.67 |
| EVA170 | 170 | 0.28 ± 0.007 | 0.45 ± 0.002 | 0.06 ± 0.003 | 0.02 ± 0.001 | 46.10 ± 2.74 |
| EVA220 | 220 | 0.25 ± 0.005 | 0.54 ± 0.007 | 0.39 ± 0.004 | 0.10 ± 0.008 | 62.45 ± 1.59 |

*3.2. Impact Attenuation Test Results*

Figures 6 and 7 shows the average force-time curves at an impact energy of 5 J and 7 J using an 8.5 kg weight following ASTM F1976, respectively. Loading rates of 55 s$^{-1}$ and 65 s$^{-1}$ for the 5 J and 7 J impact test were measured, respectively. The peak transmitted forces for the PU foams are well below those that EVA and TPU specimens transmit, thus, resulting in higher impact efficiencies. For example, at a 5 J impact energy, PU170 results in a peak force of 0.60 kN while EVA and TPU foams had an average peak force of about 0.85 kN. This represents a 28.6 % reduction in the amplitude of the force that is transmitted to the asset. Additionally, at the 7 J impact, the PU foams show a reduction in impact force of up to 36.9 % when compared to TPU and EVA foam materials. Thus, at higher energy impacts, polyurea foams display an increasing impact attenuation capability over current EVA and TPU foam technologies of similar densities while exhibiting similar deflections (Table 4).

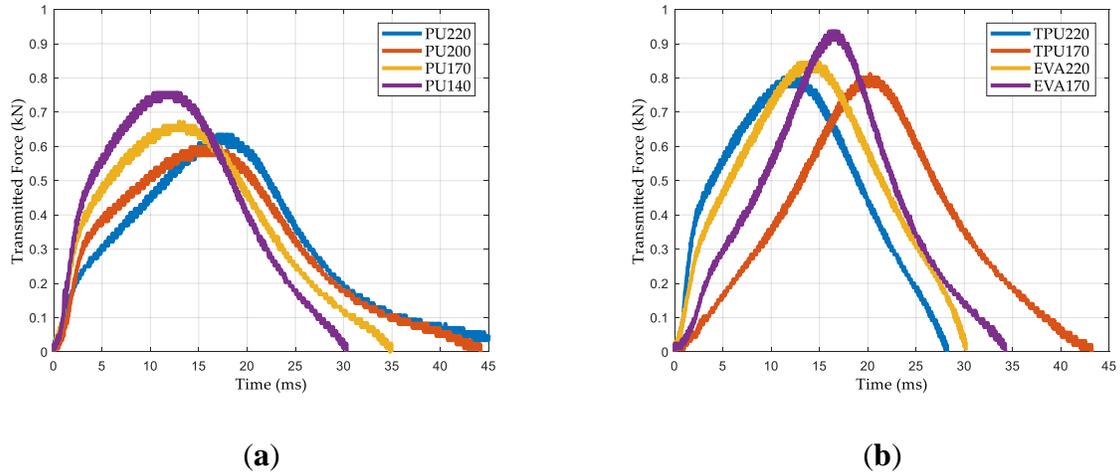

(**a**) (**b**)

**Figure 6.** Transmitted force-time history of samples impacted by 5 J energy: (**a**) PU samples; (**b**) TPU and EVA samples.

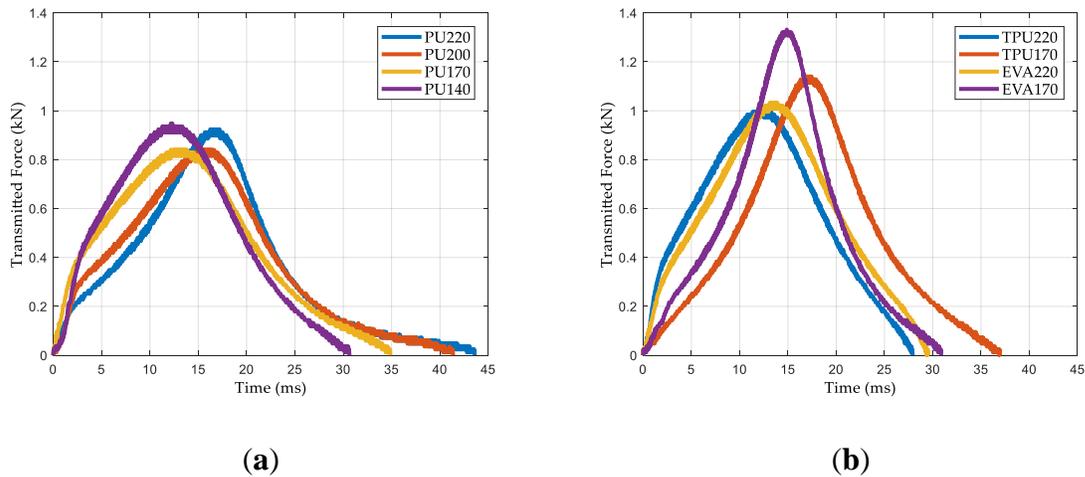

(**a**) (**b**)

**Figure 7.** Transmitted force-time history of samples impacted by 7 J energy: (**a**) PU samples; (**b**) TPU and EVA samples.

The resulting impact efficiency values of the PU samples for a 5 J impact range from 33 – 47 % while EVA and TPU foams had efficiencies ranging from 27 – 31 %. Under a 7 J impact, the PU foams have impact cushioning efficiencies ranging from 37 – 42 % while EVA and TPU foams had efficiencies ranging from 26 – 35 %. The maximum deflections in PU foams are very similar to TPU and EVA foams. However, the energy return, is lower for PU foams over both than EVA and TPU foams. PU foams are moderately resilient, while EVA and TPU foams display high resiliency. The 5 J and 7 J impact tests results are summarized in Tables 3 and 4, respectively.

**Table 3.** 5 J Impact Attenuation Test Results

| Material | Density (kg/m$^3$) | 5J Impact Test | | | |
|---|---|---|---|---|---|
| | | Peak Force (kN) | Max Disp. (mm) | Impact Efficiency (%) | Energy Return (%) |
| PU140 | 140 | 0.63 ± 0.001 | 13.92 ± 0.46 | 39.68 ± 0.06 | 25.01 ± 1.62 |
| PU170 | 170 | 0.60 ± 0.003 | 12.96 ± 0.79 | 41.67 ± 0.21 | 27.39 ± 2.84 |
| PU200 | 200 | 0.67 ± 0.002 | 11.12 ± 0.72 | 37.31 ± 0.11 | 34.28 ± 1.86 |
| PU220 | 220 | 0.76 ± 0.001 | 10.05 ± 0.66 | 32.89 ± 0.04 | 39.09 ± 1.36 |
| TPU170 | 170 | 0.80 ± 0.008 | 16.20 ± 0.73 | 31.25 ± 0.31 | 54.10 ± 2.81 |
| TPU220 | 220 | 0.82 ± 0.006 | 9.72 ± 0.59 | 30.49 ± 0.22 | 61.39 ± 2.96 |
| EVA170 | 170 | 0.94 ± 0.006 | 13.02 ± 0.21 | 26.59 ± 0.17 | 51.67 ± 1.88 |
| EVA220 | 220 | 0.84 ± 0.004 | 10.71 ± 0.39 | 29.76 ± 0.14 | 61.71 ± 1.22 |

**Table 4.** 7 J Impact Attenuation Test Results

| Material | Density (kg/m$^3$) | 7J Impact Test | | | |
|---|---|---|---|---|---|
| | | Peak Force (kN) | Max Disp. (mm) | Impact Efficiency (%) | Energy Return (%) |
| PU140 | 140 | 0.92 ± 0.009 | 12.60 ± 0.12 | 38.04 ± 0.16 | 22.08 ± 1.06 |
| PU170 | 170 | 0.84 ± 0.005 | 11.71 ± 0.36 | 41.67 ± 0.25 | 24.73 ± 2.48 |
| PU200 | 200 | 0.85 ± 0.002 | 9.83 ± 0.08 | 41.17 ± 0.10 | 25.11 ± 2.00 |
| PU220 | 220 | 0.95 ± 0.004 | 9.53 ± 0.14 | 36.84 ± 0.16 | 25.11 ± 1.96 |
| TPU170 | 170 | 1.14 ± 0.007 | 13.20 ± 0.63 | 30.70 ± 0.19 | 50.65 ± 2.81 |
| TPU220 | 220 | 1.01 ± 0.003 | 9.00 ± 0.44 | 34.65 ± 0.10 | 54.37 ± 2.21 |
| EVA170 | 170 | 1.33 ± 0.005 | 11.40 ± 0.29 | 26.31 ± 0.10 | 47.05 ± 2.71 |
| EVA220 | 220 | 1.03 ± 0.002 | 9.70 ± 0.19 | 33.98 ± 0.07 | 51.53 ± 2.61 |

It is well known that the mechanical properties of foams are derived from its cellular structure and from the polymer that forms its underlying skeleton. Therefore, the structure-property relationships for each foam system has to be identified to elucidate the energy absorbing mechanisms under impact. For a comprehensive characterization of the underlying energy dissipating mechanism in PU foams as well as in TPU and EVA foams, further analysis in determining the effect of cell size, shape, and wall thickness along with the constitutive material properties have to be conducted. Therefore, a more detailed analysis of the underlying energy dissipating mechanism is currently under investigation and will be reported in a future communication.

## 4. Conclusions

The cushioning characteristics and impact attenuation of low density polyurea (PU) foams (140 - 220 kg/m$^3$) were presented. The stress-strain characteristics, cushioning efficiency, and energy return (resilience) were measured at a quasi-static strain rate. PU foams displayed higher energy absorption capacity and higher cushioning efficiencies than widely used EVA and TPU foams. In addition, the impact attenuation of the polyurea foams at 5 J and 7 J impact energies were investigated according to the American Society for Testing and Materials (ASTM) test methods F1976. Under impact testing, PU foams resulted in a reduction of up to 28.6 % and 36.9 % in peak transmitted impact forces when compared to EVA and TPU foams of the same thickness (20 mm) at 5 J and 7 J, respectively. These properties should allow these materials to have a wide range of cushioning/impact applications, especially in protective body and headgear systems.

**Acknowledgments:** This work was supported by the Office of Naval Research under grant number N0008 for which we are grateful to Dr. Roshdy Barsoum of the agency.

**Conflict of Interest:** The authors declare that they have no conflict of interest.

**References**

1. Gibson, L.J.; Ashby, M.F. *Cellular Solids: Structure and Properties*, 2nd ed.; Cambridge University Press, Cambridge, UK, 1999.

2. Mills, N. *Polymer Foams Handbook: Engineering and Biomechanics Applications and Design Guide*, Butterworth-Heinemann 2007.

3. K. C. Rusch, J. Appl. Polym. Sci. 13, 2297 (1969).

4. K. C. Rusch, J. Appl. Polym. Sci. 14, 1433 (1970).

5. M. Avalle, G. Belingardi, R. Montanini, International Journal of Impact Engineering 25, 455 (2001).

6. J. Miltz, O. Ramon, Polym. Eng. Sci. 30, 129 (1990).

7. V. Gupta, G. Youssef, Exp. Mech. 2014, 54, 1133 (2014).

8. B. J. Ramirez, O. T. Kingstedt, R. Crum, C. Gamez, V. Gupta, Journal of Applied Physics 121, 22 (2017).

9. B. J. Ramirez, V. Gupta, Mater. Des. 137, 298 (2018).

10. B. J. Ramirez, U. Misra, V. Gupta, Mechanics of Materials 127, 39 (2018).

11. ASTM Standard F1976 (2006) Standard test method for impact attenuating of athletic shoe cushioning systems and materials. West Conshohocken: ASTM International.

12. V. Gupta, R. Crum, C. Gámez, B. Ramirez, N. Le, G. Youssef, J. Citron, A. Kim, A. Jain, U. Misra. Adhesive and Ultrahigh Strain Rate Properties of Polyurea Under Tension, Tension/Shear, and Pressure/Shear Loadings with Applications to Multilayer Armors. In *Elastomeric Polymers with High Rate Sensitivity: Applications in Blast, Shockwave, and Penetration Mechanics*, 1st, ed.; Barsoum, R.; Elsevier: Oxford, UK, 2015; Volume 1, pp. 71-92, 978-0-323-35400-4. http://dx.doi.org/10.1016/B978-0-323-35400-4.00003-9

13. B. J. Ramirez, Dissertation, Manufacturing and Characterization of Temperature-Stable, Novel, Viscoelastic Polyurea Based Foams for Impact Management. University of California, Los Angeles (2017). https://escholarship.org/uc/item/5qk268pn

14. B. J. Ramirez, V. Gupta, International Journal of Mechanical Sciences (Accepted).